\documentstyle [12pt] {article}
\newcommand{\cs}[3]{{{#3} \brace {#1 #2}}}

\newcommand{\h}[1]{\mathop{\lambda}\limits_{#1}\ \!\!\!}

\newcommand{\edf}{\ {\mathop{=}\limits^{\rm def}}\ }

\newcommand{\al}{\alpha}
\newcommand{\be}{\beta}
\newcommand{\m}{\mu}
\newcommand{\n}{\nu}
\newcommand{\s}{\sigma}
\newcommand{\g}{\gamma}

\newcommand{\de}{\delta}

\begin{document}
\title{\bf Parameterized Absolute Parallelism: A Geometry for Physical Applications}

\author{{\bf M. I. Wanas}\\
\normalsize Astronomy Department, Faculty of Science,
Cairo university {\bf Egypt}\\
e. mail wanas@frcu.eun.eg}

\maketitle
\begin{abstract}

Absolute parallelism (AP) geometry is frequently used for physical
applications. Although it is wider than the Riemannian geometry,
it has two main defects. The first is that its path equation does
not represent physical trajectories of any test particle. The
second is the identical vanishing of its curvature tensor. The
present work shows that parameterizing this geometry would solve
the two problems. Furthermore, the resulting parameterized
(AP)-structure is more general than both the conventional
(AP)-structure and the Riemannian structure. Also, it is shown
that it can be reduced to one or the other, of these two geometric
structures, in some special cases. The structure obtained is more
appropriate for physical applications, especially in constructing
field theories gauging gravity.
\end{abstract}

\newpage

\section{Introduction}

Geometrization of physics is an important philosophy introduced by
Albert Einstein at the beginning of the twentieth century. It
represents, together with the quantization philosophy, the main
two ideas of fundamental physics of the century. It proved to be
very successful in describing and interpreting gravitational
interaction. During the first half of the century, and until the
end of his life, Einstein tried to include other interactions in
his scheme, especially the electromagnetic one, in a series of
attempts known as {\it Unified Field Theories}. Unfortunately, all
these attempts were not successful or incomplete.

On the other hand, in the second half of the century, several
attempts have been done to unify the four interactions known to
physicists, so far, using the quantization philosophy. A partial
success is achieved in the trial known as {\it The Standard Model}
of Glashow, Weinberg and Salam (cf. [1]), which unifies weak and
electromagnetic interactions. The success of this trial encouraged
authors to rush in this direction. Another trial, of less success,
is done to include the strong interaction together with weak and
electromagnetic interactions, known as {\it Grand Unified
Theories} (cf. [1]]. Extensive efforts have been done to include
gravity in this scheme. The problem is that, it seems very
difficult or probably impossible to unify gravity with the other
interactions using the conventional quantization scheme. It is not
convincing that the continuation of the attempts in this direction
will bring us more nearer to the solution of the problem.

The situation now is either to suggest a third philosophy or to
return back and reexamine carefully the geometrization scheme used
to construct general relativity (GR). In the present work, I am
going to follow the easier way, i.e. to reexamine the
geometrization scheme. Consequently, I am going to suggest a
modified version of {\it Absolute Parallelism} (AP) geometry more
appropriate for physical applications than the conventional
version.

\section{Scheme of Geometrization}
Geometrization philosophy can be summarized in the following
statement: {\it "to understand nature one should start with
geometry and end with physics"}. The theory of GR is a successful
application of this philosophy. The application of this philosophy
implies the following scheme. It contains the following
criteria:\\ 1- A certain geometry is chosen to represent a model
for nature. It should be wide enough to accommodate all required
physical quantities.\\ 2- Building blocks in this geometry should
be of one type.\\ 3- Spaces of the chosen geometry should be
affinely connected in order to guarantee general covariance of
mathematical expressions.\\ 4- A one-to-one correspondence is set
between geometric objects and physical quantities.\\ 5- Identities
in the chosen geometry represent laws of nature.\\ 6- Curves or
paths in this geometry represent trajectories of test particles.\\
7- A theory in which matter is not represented by a
phenomenological material-energy tensor is much preferable [2].\\
8- A theory in which different interactions do not enter as
logically distinct structures would be much preferable [3].\\

To illustrate the above criteria, we use the most successful
application of the geometrization scheme, GR in free space. For
the first criterion, the geometry chosen for this application is
the Riemannian geometry. It is just sufficient for the description
of gravitational interaction. The building block of the geometry
is the metric tensor only. All other geometric objects are
constructed using this tensor. The affine connection defined is
Christoffel symbol, and consequently Riemannian spaces are
affinely connected. The Bianchi identity is used to represent
conservation, as a law of nature. Geodesic and null-geodesic paths
are used to represent, successfully, trajectories of test
particles and of photons respectively. There is no
phenomenological material-energy tensor imposed on the theory, and
the sources of the gravitational field arise as constants of
integration of the field equations. The eighth criterion is not
applicable in this case since the theory is for gravitational
interaction only. One can classify theories constructed using the
above scheme, in which the eight criteria are satisfied, as {\it
"Pure Geometric Field Theories"}.

Einstein spent the last thirty years of his life trying to
construct geometric field theories, in two main series of attempts
to unify gravity with electromagnetism. These attempts are known
in the literature as {\it "Unified Field Theories"}. In the first
series of these attempts, that we call {\it "Einstein's Absolute
Parallelism Theory"} (EAP), he used the geometry of absolute
parallelism [4] to construct a theory unifying gravity and
electromagnetism. For the same reason he started the second series
[3], known as {\it "Einstein's Non-Symmetric Theory"} (ENS), using
another type of non-symmetric geometry in which he dropped the
symmetry conditions from the metric tensor and from the affine
connection. Table 1 summarizes the application of the
geometrization scheme, mentioned above, in the two series of
attempts compared to GR.
\newpage

\begin{center}
Table 1: Are the criteria of the geometrization scheme satisfied?
\vspace{.3cm}
\begin{tabular}{|c|c|c|c|} \hline
 & & & \\
Criterion & GR & EAP & ENS  \\
\hline
 & & & \\
1-Geometry   & Yes, Riemannian & Yes, AP & Yes, Non-symmetric \\
 & & &  \\
\hline
 & & &    \\
2-Building Blocks & Yes, $ g_{(\m\n)}$ & Yes, $\h{i}^\m$ & (?), $
g_{\m\n}$ \\
 & &  & \\
\hline
 & & & \\
3-Affine Connection  & Yes, $ g_{\m\n;\al}=0$ & Yes, $\h{i}_{{\stackrel{\m}{+}} | \n} = 0$ & Yes, $g_{{{\stackrel{\m}{+}}{\stackrel{\n}{-}}} | \al}=0$  \\
 & & & \\
\hline
 & & &  \\
4-Correspondence & Yes & Yes & Yes \\
 & & &  \\
\hline
 & & & \\
5-Identities &  Bianchi & (?) & Generalized Bianchi \\
 & & &  \\
\hline
 & & & \\
6- Paths and Motion & Yes, Geodesic & No & No, in general. \\
 &  & &  \\
\hline
 & & & \\
7-Exclusion of & & & \\
Phenomenological & Yes, Free space & Yes & Yes \\
Objects  & & & \\
 & & &  \\
\hline
 & & & \\
8-Unique Structure & Yes & Yes & No  \\
 & & & \\
\hline
\end{tabular}
\end{center}

\newpage
From the above table it is clear that all the criteria of the geometrization
scheme are satisfied in the case of GR in free space.

The EAP-theory [4] has been found to be unsatisfactory since it
did not produce the Schwarzschild field, in the case of spherical
symmetry, when the electromagnetic field vanished. Also, it has
been found, as shown in Table 1, that autoparallel paths do not
represent trajectories of any known particles. Furthermore, some
authors (e.g. [5], [6]) claim that the AP-geometry is a flat one,
since a curvature tensor, defined in a certain way, vanishes. The
last two objections indicate, in my opinion, the incompleteness of
the version of the AP-geometry used in constructing the theory. I
am going to stress on these objections, and to give them more
attention in the next Section.

For the ENS-theory [3], most of the criteria are satisfied. one objection is
that the metric tensor is non-symmetric. It is made of two parts: one is
symmetric and the other is skew. The skew part has no contribution to the
quadratic expression giving the metric of space. The term containing the skew
part, in this expression, vanishes identically. So the geometry may be considered
as a symmetric one to which one can add any second order skew tensor. Then if
the symmetric part is related to the gravitational field, as usually done, and
the skew part to the electromagnetic field, so the two fields enter as logically
two distinct structures. This violates the 8th criteria. Another objection, of
less importance, is that the non-symmetric connection in this geometry is defined
using a metricity condition:

$$
 g_{{{\stackrel{\m}{+}}{\stackrel{\n}{-}}} | \al}=0, \eqno{(1)}
$$ whose solution, in this case, gives a complicated expression
which makes subsequent calculations tedious. A third objection,
against this geometry, is that paths of this geometry do not, in
general, represent trajectories of test particles except for some
restricted cases (cf. [7]).

Several authors tried to construct field theories using the above
mentioned geometries. Although their attempts have satisfied many
of the criteria mentioned above, yet these attempts violate
others. Some of these attempts are listed in Table 2, in an
ascending order of the year of publication in the first column.
The second column of this table gives the aim of the trial. The
headings of the last eight columns give the criterion number, in
the scheme of geometrization given above. The table shows how many
of these criteria are satisfied by the listed attempts.

\newpage
\begin{center}
Table 2: Confrontation between Geometrization attempts and
Criteria \vspace{1.3cm}
\begin{tabular}{|c|c|c|c|c|c|c|c|c|c|} \hline
& & & & & & & & & \\ Trial (year) & Aim & 1 & 2 & 3 & 4 & 5 & 6 &
7 & 8 \\ & & & & & & & & & \\ \hline & & & & & & & & & \\
Levi-Civita(1950) [8] & Simplification of EAP & $\surd$ & $\surd$
& $\surd$ & $\surd$ & x & x & x & $\surd$ \\ & & & & & & & & & \\
\hline & & & & & & & & & \\ Mikhail(1952) [9] & Creation of Matter
& $\surd$ & $\surd$ & $\surd$ & x & x & x & x & $\surd$ \\ & & & &
& & & & & \\ \hline & & & & & & & & & \\ M\o ller (1961) [10] &
Energy Localization & $\surd$ & $\surd$ & $\surd$ & x & $\surd$ &
x & x & $\surd$ \\ & & & & & & & & & \\ \hline & & & & & & & & &
\\ Mikhail(1964) [11] & Unification & $\surd$ & $\surd$ & $\surd$
& x & x & x & $\surd$ & $\surd$ \\ & & & & & & & & & \\ \hline & &
& & & & & & & \\ Mikhail$\&$Wanas(1977) [12] & Unification &
$\surd$ & $\surd$ & $\surd$ & $\surd$ & $\surd$ &  x & $\surd$ &
$\surd$
\\ & & & & & & & & & \\ \hline & & & & & & & & & \\ M\o ller
(1978) [13] & Singularity Free & $\surd$ & $\surd$ & $\surd$ & x &
$\surd$ & x & x & $\surd$ \\ & & & & & & & & & \\ \hline & & & & &
& & & & \\ Hayashi$\&$Shirafuji(1979) [5] & Microscopic Gauge
Gravity & $\surd$ & $\surd$ & $\surd$ & ? & ? & ? & x & $\surd$ \\
& & & & & & & & & \\ \hline & & & & & & & & & \\ Hammond(1990)
[14] & Dynamical Torsion & $\surd$ & x & $\surd$ & $\surd$ & ? & x
& x & x \\ & & & & & & & & & \\ \hline & & & & & & & & & \\
Hammond(1995) [15] & Dirac Field $\&$ GR & $\surd$ & x & $\surd$ &
$\surd$ & x & ? & x & x \\ & & & & & & & & & \\ \hline & & & & & &
& & & \\ Moffat(1995) [16] & Non-Symmetric Gravity & $\surd$ &
$\surd$ & $\surd$ & ? & $\surd$ & ? & x & $\surd$ \\ & & & & & & &
& & \\ \hline & & & & & & & & & \\ Hammond(1998) [17] &
Geometrization of Strings & $\surd$ & x & $\surd$ & $\surd$ & x &
? & x & x \\ & & & & & & & & & \\ \hline
\end{tabular}
\end{center}
$\surd =$ Criterion satisfied \\
x = Criterion not satisfied \\
? = Not clear \\
\section{Why Absolute Parallelism?}
Before giving an answer to this question, I will review briefly
the fundamental bases of the conventional AP-geometry. An AP-
space is an n-dimensional space each point of which is labeled by
a set of variables $x^\n (\n =1, 2, 3, ..., n)$. At each point we
define n-contravariant vectors $\h{i}^\m (\m = 1, 2, 3, ..., n$
stands for the coordinate components and $i = 1, 2, 3, . .., n$
stands for the vector number). We are going to use Latin indices
for vector numbers and Greek indices for the coordinate
components. These vectors are subject to the condition: $$
\h{i}_{{\stackrel{\m}{+}} | \n} = 0, \eqno{(2)} $$ which is the
absolute parallelism condition. One can define a set of n-
covariant vectors, conjugate to $\h{i}^\m$ such that,

$$ \h{i}^\m \h{j}_\m=\de_{ij}, \eqno{(3)} $$

$$
 \h{i}^\m \h{i}_\n=\de^\m_{\ \n}. \eqno{(4)}
 $$
Summation convention will be applied to both types of indices.
Using these vectors we can define the following 2\underline {nd}
order symmetric tensor: $$ g_{\n\m} \edf \h{i}_\n \h{i}_\m,
\eqno{(5)} $$

$$ g^{\n\m} \edf \h{i}^\n \h{i}^\m. \eqno{(6)} $$ One can then use
$g_{\m\n}$ to play the role of the metric tensor of the Riemannian
geometry, and so it can be used with its conjugate to lower and
raise Greek indices respectively.\\

\underline{Affine Connections}: In this geometry we have more than
one affine connection. A non-symmetric connection is immediately
obtained as a solution of the AP-condition given by (2), $$
\Gamma^\al_{.\ \m\n}\edf \h{i}^\al\h{i}_{\m,\n}. \eqno{(7)} $$
Also its dual behaves as an affine connection under the group of
general coordinate transformations. This connection is written in
the form: $$ {\hat{\Gamma}}^\al_{.\ \m\n} \edf {\Gamma}^\al_{.\
\n\m} . \eqno{(8)} $$ Since (7) is non-symmetric, its symmetric
part, $$ \Gamma^\al_{.\ (\m \n)}\edf {1 \over 2}{(\Gamma^\al_{.\
\m \n} + \Gamma^\al_{.\ \n \m})} , \eqno{(9)} $$ behaves as an
affine connection. Now since (5) is defined as a metric tensor,
then as a consequence of a metricity condition, $$ g_{{\m \n}\ ;
\s} = 0 , \eqno{(10)} $$ we can define Christoffel symbol in the
usual manner, $$ \cs{\m}{\n}{\al} \edf {1 \over 2} g^{\al \s}
(g_{\m \s , \n} + g_{\s \n , \m} - g_{\m \n , \s}) . \eqno{(11)}
$$ It is worth of mention that the symmetric part of the
non-symmetric connection (9) is different from Christoffel symbol.
We can define using this connection a third order skew tensor,
known as the torsion tensor,

$$ \Lambda^{\al}_{.\ \m\n}\edf\Gamma^{.\al}_{\
\m\n}-\Gamma^{\al}_{.\ \n\m}=-\Lambda^{\al}_{.\ \n\m} .
\eqno{(12)}$$ Consequently, one can define another non-symmetric
connection by adding (11) to (12), $$ {\Omega}^\alpha _{.~\mu \nu}
\edf \cs{\m}{\n}{\al}\ + \Lambda^{\al}_{.~\m \n}, \eqno{(13)}$$
from which it is clear that its symmetric part is Christoffel
symbol. The dual of (13) is also an affine connection, given by $$
{\hat{\Omega}}^\alpha _{.~\mu \nu} \edf {\Omega}^\alpha _{.~\nu
\mu}, \eqno{(14)} $$ So, in the AP-geometry one can define at
least six quantities which behave as affine connections under the
effect of the group of general coordinate transformations. Some of
these quantities are symmetric (9) and (11), while others are
non-symmetric. (7), (8), (13) and (14).\\

\underline{Absolute Derivatives}: Using the above defined affine
connections, one can define the following absolute derivatives: $$
A^{\m}_{+|~ \n} = A^{\m}_{~, \n} + A^{\al}~{\Gamma}^{\m}_{.~\al
\n},  \eqno{(15)} $$ $$ A^{\m}_{-| \ \n} = A^{\m}_{\ , \n} +
A^{\al} {\hat{\Gamma}}^{\m}_{. \al \n},  \eqno{(16)}$$ $$
A^{\m}_{0 |~ \n} = A^{\m}_{~, \n} + A^{\al}~{\Gamma}^{\m} _{.~(\al
\n)}, \eqno{(17)}$$ $$ A^{\m}_{+||~ \n} = A^{\m}_{~, \n} +
A^{\al}~{\Omega}^\mu _{.~\alpha \nu}, \eqno{(18)} $$ $$
A^{\m}_{-||~ \n} = A^{\m}_{~, \n} + A^{\al}~{\hat{\Omega}}^\mu
_{.~\alpha \nu} , \eqno{(19)} $$ $$ A^{\m}_{0 || \n} = A^{\m}_{~,
\n} + A^{\al} {\Omega}^\mu _{. (\alpha \nu)}, \eqno{(20)}$$ where
$A^{\m }$ is any contravariant vector. The last derivative is
equivalent to the conventional covariant derivative since the
symmetric part of the connection ${\Omega^\mu_{.\ \alpha \nu}}$ is
Christoffel symbol, as stated above.\\ \underline{Basic Vector}:
We can define a third order tensor, called the contortion, as $$
\g^\al_{.\ \m\n}\edf\h{i}^\al \h{i}_{\m ;\n}. \eqno{(21)}$$

It can be easily shown that $$
 \Gamma^\al_{.\ \m\n}=\cs{\m}{\n}{\al}+\gamma^\al_{.\ \m\n}, \eqno{(22)}
$$ $$ \Lambda^{\al}_{.\ \m\n}\edf\g^{\al}_{.\ \m\n} - \g^{\al}_{.\
\n\m}=-\Lambda^{\al}_{.\ \n\m}, \eqno{(23)} $$ $$ \g^{\al}_{. \m
\n} {\ } \edf \frac{1}{2}(\Lambda^{\al}_{.\m \n} - {\Lambda}_{\m
.\n }^{~\al} - {\Lambda}_{\n .\m}^{~\al}). \eqno{(24)}$$
Contracting we can define a covariant basic vector, $$
C_{\m}\edf\Lambda^\al_{.\ \m\al}=\g^\al_{.\ \m\al}. \eqno{(25)}$$
\\
\underline{Path Equation}: Historically, it is known that the only
path defined in the AP-geometry is the autoparallel path given by:
$$ {\frac{dA^\m}{dS}} + \Gamma^{\m}_{.\al \be}\ A^\al A^\be = 0 .
\eqno{(26)}$$ One of the reasons for which EAP-theory and the
AP-geometry were neglected for about twenty years (form
Robertson's paper [18] in 1932 to Mikhail's Ph.D. [9] in 1952) is
that this path does not represent trajectories of any known
particles. However, recently [19] it is shown that the AP-geometry
admits other paths, whose equations can be written in the form: $$
{\frac{dV^\m}{dS^+}} + \{^{\m}_{\al\be}\} V^\al V^\be = -
\Lambda^{~ ~ ~ ~ \m}_{(\al \be) .} ~~~V^\al V^\be,  \eqno{(27a)}
\\ $$ $$ {\frac{dW^\m}{dS^0}} + \{^{\m}_{\al\be}\} W^\al W^\be = -
{\frac{1}{2}} \Lambda^{~ ~ ~ ~ \m}_{(\al \be) .}~~~ W^\al W^\be,
\eqno{(27b)}\\ $$ $$ {\frac{dU^\m}{dS^-}} + \{^{\m}_{\al\be}\}
U^\al U^\be = 0, \eqno{(27c)} $$ where$S^{+}$, $S^{0}$ and $S^{-}$
are the evolution parameters characterizing the three paths
respectively
; and  $V^{\alpha}, W^{\alpha}$ and$U^{\alpha}$ are the tangents
to the corresponding paths . We will discuss this set of equations
at the end of this Section and in the next one.\\

\underline{Curvature Tensors}: In general, there are at least two
different methods to define curvature tensors in any affinely
connected geometry.

{\it The First Method}: In this method we simply replace
Christoffel symbol, in the definition of Riemann-Christoffel
tensor, by the non-symmetric connection (7), then we get:

$$ B^{\al}_{.\m \n \s} {\ }  = {\ } \Gamma^{\al}_{.\m  \s, \n} -
\Gamma^{\al}_{.\m \n, \s} + \Gamma^{\al}_{\epsilon \n} \Gamma^
{\epsilon}_{. \m \s} - \Gamma^{\al}_{. \epsilon \s}
\Gamma^{\epsilon}_{. \m \n}. \eqno{(28)}$$

Unfortunately the curvature tensor, defined in this way, vanishes
identically because of the AP-condition (2). Some authors believe
that Ap-space is a flat space because of the vanishing of this
curvature. There is no convincing reason for which authors
choosing a certain connection, (7), while neglecting others, then
claiming that the space is flat! I will show, in the next Section,
that the space is \underline{not} flat and (28)can be considered
as one of the advantages of the AP-geometry.

{\it The Second Method}: An alternative method, for defining
curvature tensors, is to consider it as a measure of
non-commutation of the absolute derivatives given above. To make
calculations easier, and the geometry self consistent, it is
better to use the contravariant components of the tetrad vectors
in the definition of the following curvature tensors [12], [20],
$$ \h{i}^{\stackrel{\m}{+}} | {\n \s}  - \h{i}^{\stackrel{\m}{+}}
| {\s \n}   = \h{i}^\al B^\m_{. \al \n \s}, \eqno{(29)}$$ $$
\h{i}^{\stackrel{\m}{-}} | {\n \s}  - \h{i}^{\stackrel{\m}{-}} |
{\s \n}   = \h{i}^\al L^\m_{. \al \n \s}, \eqno{(30)}$$ $$
\h{i}^{\stackrel{\m}{0}} | {\n \s}  - \h{i}^{\stackrel{\m}{0}} |
{\s \n}   = \h{i}^\al N^\m_{. \al \n \s}, \eqno{(31)}$$ $$
\h{i}^{\stackrel{\m}{+}} || {\n \s}  - \h{i}^{\stackrel{\m}{+}} ||
{\s \n}   = \h{i}^\al M^\m_{. \al \n \s}, \eqno{(32)}$$ $$
\h{i}^{\stackrel{\m}{-}} || {\n \s}  - \h{i}^{\stackrel{\m}{-}} ||
{\s \n}   = \h{i}^\al K^\m_{. \al \n \s}, \eqno{(33)}$$ $$
\h{i}^{\stackrel{\m}{0}} || {\n \s}  - \h{i}^{\stackrel{\m}{0}} ||
{\s \n}   = \h{i}^\al R^\m_{. \al \n \s} . \eqno{(34)}$$ Here
again the curvature defined by (29) vanishes identically because
of the AP- condition. The tensor defined by (34) is the
Riemann-Christoffel curvature tensor of the associated Riemannian
space, since the symmetric part of the connection ${\Omega}^\mu
_{.~\alpha \nu}$ is Christoffel symbol as stated above.

It is worth of mention that the two methods, in the case of
Riemannian geometry, give identical results, since the affine
connection in this case is unique in this geometry.

Now, we answer the question given in the title of the present
Section: Why absolute parallelism? Recalling that the structure of
any AP-space is defined completely, in four dimensions,  by a
tetrad vector field subject to the condition (2), we can summarize
the advantages of using the AP-geometry, in the geometrization
scheme, in the following points:\\ 1- Calculations using this
geometry are more easier than those using Einstein's non-symmetric
geometry. All tensors of different orders, and affine connections,
are explicitly defined by relatively simple expressions.\\ 2- The
AP-geometry is more wider than the Riemannian one. It admits
tensors of third orders (21)$\&$(23), a basic vector (25) and a
number of second order skew and symmetric tensors. It also admits
more than one affine connection, some of which are symmetric and
others are non- symmetric.\\ 3- This type of geometry admits
non-vanishing torsion (12). Recently, it is shown that torsion is
necessary to couple Dirac field to gravity [15]. Also, it is
suggested that torsion is necessary to geometrize strings
[17].Furthermore, it appears that gauge formulation of gravity
needs non-vanishing torsion [15].\\ 4- Tetrads, defining the
structure of the AP-space, are used as fundamental variables in
attempts to quantize gravity [15].\\ 5- A metric tensor is
defined, in the AP-spaces, whenever needed. In other words, there
is always a Riemannian space associated with any AP-space. This
facilitates comparison between any field theory written in the
AP-space and GR.\\ 6- The use of the AP-geometry helped in solving
some of the problems of GR, e.g. the problem of localization of
gravitational energy [10].\\ 7- AP-geometry admits number of path
equations (27)[19], in which the effect of the torsion appears
explicitly.\\ 8- Quantum features are recently discovered in such
type of geometry [22].\\ 9- The AP-condition (2) is necessary to
describe the dynamics of spinning particles [23].\\ 10- Using the
tetrad vectors, one can always associate a set of scalars with
each tensor defined in the AP-geometry (cf. [24]).\\

\section{Physical Needs for Parameterizing AP-Geometry}

We have at least two convincing physical reasons for
parameterizing this type of geometry. These two reasons are:\\

{\it The First Reason:} Let us examine the structure of the
curvature tensor given by (28). As stated before, this tensor
vanishes identically because of the AP-condition (2). This tensor
can be written in the form, $$ B^\al_{.\ \m \n \s} \edf R^\al_{.\
\m \n \s} + Q^\al_{.\ \m \n \s}, \eqno{(35)} $$ where $ R^\al_{.\
\m \n \s} $ is the Riemann-Christoffel curvature tensor, of the
associated Riemannian space, given by, $$ R^\al_{.\ \m \n \s} \edf
{\cs{\m}{\s}{\al}}, \n - {\cs{\m}{\n}{\al}}, \s + \cs{\m}{\s}{\be}
\cs{\be}{\n}{\al} - \cs{\m}{\n}{\be} \cs{\be}{\s}{\al},
\eqno{(36)}$$ and $$ Q^\al_{.\ \m \n \s} \edf
\g^{\stackrel{\al}{+}}_{{.\ {\stackrel{\m}{+}}{\stackrel{\n}{+}}}
| \s} - \g^{\stackrel{\al}{+}}_{.\
{{\stackrel{\m}{+}}{\stackrel{\s}{-}}} | \n} + \g^{\be}_{. \m \s}
{\ } \g^{\al}_{. \be \n} {\ } - \g^{\be}_{. \m \n} {\ }
\g^{\al}_{. \be \s}.\eqno{(37)}$$ It is clear from (36) that
$R^\al_{.\ \m \n \s}$ is made from Christoffel symbols only. Also
from (37) we can see that $Q^\al_{.\ \m \n \s}$ is made from the
contortion (or the torsion via (24)) only. Some authors believe
that $R^\al_{.\ \m \n \s}$ and $Q^\al_{.\ \m \n \s}$ are
equivalent (cf.[6]). Others consider $Q^\al_{.\ \m \n \s}$ as
giving an alternative definition of $R^\al_{.\ \m \n \s}$ Let us
examine these two tensors from a different point of view. It is
well known that Christoffel symbol is related, in applications, to
the gravitational field. So, its existence in (36) indicates that
gravity is responsible for the curvature of space-time. The
identical vanishing of the curvature $B^\al_{.\ \m \n \s}$ may
indicate that there is another physical interaction (anti-gravity,
say) which is related to the contortion (or the torsion) and is
represented by the tensor $Q^\al_{.\ \m \n \s}$. This interaction
balances the effect of gravity in such a way that the total effect
vanishes. If so, it is better to call the tensor $Q^\al_{.\ \m \n
\s}$ \underline{\it "The Curvature Inverse of Riemann-Christoffel
Tensor"}. But since gravity dominates our observable universe,
which indicates that $ R^\al_{.\ \m \n \s} $ is more effective
than $Q^\al_{.\ \m \n \s}$, thus one has to parameterize torsion
terms in AP- expressions.

{\it The Second Reason:} The set of equations (27) represents
paths, admitted by the AP-geometry, that are different from those
of the Riemannian geometry. Although the first equation of this
set is similar to the geodesic (or null-geodesic) equation, the
other two equations cannot be reduced to the geodesic one, unless
the torsion vanishes. It has been shown that the vanishing of the
torsion of the AP-space will reduce the space to a flat one [20].
This is because all curvature tensors, defined in the previous
Section, can be written in terms of the torsion (or contortion)
tensor. Consequently, all of these tensors vanish as a result of
the vanishing of the torsion. So, what are the trajectories of
particles that can be represented by these two equations? Clearly
there are no particles that move along these paths. The reason is
that the effect of the Christoffel symbol term, in these
equations, is of the same order of magnitude as the effect of the
torsion term. So, for these equations to represent physical
trajectories, the torsion term in the equations should be
parameterized, in order to reduce its effect [25].

\section{Parameterized AP-Geometry}
As it is shown in the previous Section, the two reasons for which
we parameterize the  geometry are the vanishing of the curvature
tensor (28) and the problem of the physical meaning of the set of
path equations (27). As stated in the previous Section, the common
factors between these two features are the affine connections. So,
it is necessary to start parameterizing these connections first.

\underline{\it Parameterized Connection}: One way to parameterize
the AP-geometry is to define a general affine connection by
linearly combining the affine connections defined in the geometry.
In doing so, we get after some manipulations [25]: $$
\nabla^\m_{.\al \be} = a_1  {\cs{\al}{\be}{\m}} + (a_2 - a_3)
\Gamma^{\m}_{. \al \be} - (a_3 + a_4)\Lambda^{\m}_{. \al \be}{\
}{\ }{\ }, \eqno{(38)} $$ where $a_1 , a_2, a_3$ and $a_4$ are
parameters. It can be easily shown that $\nabla^\mu_{.\alpha
\beta}$ transforms as an affine connection under the group of
general coordinate transformations.It is clear that this
parameterized connection is non-symmetric.

\underline{\it Parameterized Absolute derivatives}: If we
characterize absolute derivatives, using the connection (38), by a
treble stroke, then we can define the following derivatives: $$
A^{\m}_{+|||~ \n} \edf A^{\m}_{~, \n} + A^{\al}\nabla^\m _{.\al
\n}, \eqno{(39)} $$ $$ A^{\m}_{-|||~ \n} \edf A^{\m}_{~, \n} +
A^{\al}\nabla ^\m_{.\n \al}, \eqno{(40)} $$ $$ A^{\m}_{0|||~ \n}
\edf A^{\m}_{~, \n} + A^{\al}\nabla ^\m _{.(\al \n)}, \eqno{(41)}
$$ where $A^\m$ is any arbitrary vector. If we need metricity,
using the parameterized connection, to be preserved i.e., $$
g_{\stackrel{\m}{+} \stackrel{\n}{+} |||\sigma} = 0, \eqno{(42)}
$$ then one should take, $$
  a+b=1, \eqno{(43)}
$$ where  $a=a_1$, $b=a_2+a_4$, ($a_3 = -a_4$) are two parameters.
Now, due to (43), we have only one parameter. In this case the
general affine connection (38) can be written in the form: $$
\nabla^\al_{. \m \n} =  {\cs{\m}{\n}{\al}} + b \g^\al_{. \m \n} .
\eqno{(44)}$$ It is clear from this equation that we have
parameterized the contortion (or equivalently the torsion) term in
a general connection of the AP-geometry. Now we will explore the
consequences of this parameterization.

\underline{\it Parameterized Path Equation}: Using the
parameterized connection
 (44) and following the same approach followed before to get the set (27),
 we can get the following parameterized path equation admitted by the geometry [25],
$$ {{\frac{dZ^\m}{d\tau}} + \cs{\n}{\s}{\m}\ Z^\n Z^\s = - b{\ }
\Lambda_{(\n \s)}.^\m~~  Z^\n Z^\s}, \eqno{(45)}$$ where $Z^\m$
$(\edf \frac{dx^\m}{d\tau})$ is the tangent to the path and $\tau$
is the evolution parameter along it.
 This equation replaces the set given by (27). Discussion of this result is
 given in the next Section.

\underline{\it Parameterized Curvature Tensors}: Using the first
method, given in Section 3, for defining curvature tensors we can
define the following tensor, $$ {\hat {B}}^\al_{\ \m \n \s} \edf
\nabla^\al_{. \m \s , \n} - \nabla^\al_{. \m \n , \s} +
\nabla^\be_{. \m \s} \nabla^\al_{. \be \n} - \nabla^\be_{. \m \n}
\nabla^\al_{. \be \s}. \eqno{(46)}$$ Using the definition of
$\nabla^\be_{. \m \n} $ given by (38) and applying the metricity
condition (43), then we can write, $$ {\hat {B}}^\al_{.\ \m \n \s}
= R^\al_{.\ \m \n \s} + b {\hat {Q}}^\al_{.\ \m \n \s} ,
\eqno{(47)}$$ where $$ {\hat {Q}}^\al_{.\ \m \n \s} \edf
\g^{\stackrel{\al}{+}}_{.\ {{\stackrel{\m}{+}}{\stackrel{\n}{+}}}
| \s} - \g^{\stackrel{\al}{+}}_{.\
{{\stackrel{\m}{+}}{\stackrel{\s}{-}}} | \n}  + b (\g^{\be}_{. \m
\s} {\ } \g^{\al}_{. \be \n} {\ } - \g^{\be}_{. \m \n} {\ }
\g^{\al}_{. \be \s} {\ }).  \eqno{(48)}$$ It is clear that the
tensor ${\hat {B}}^\al_{.\ \m \n \s}$ is a parameterized
replacement of the tensor $B^\al_{.\ \m \n \s}$ given in Section
3. But here ${\hat {B}}^\al_{.\ \m \n \s}$ is, in general,
non-vanishing.

Using the second method, given in Section 3, for defining
curvature tensors we get the following tensors,

$$ \h{i}^{\stackrel{\m}{+}} ||| {\n \s}  -
\h{i}^{\stackrel{\m}{+}} ||| {\s \n}   = \h{i}^\al W^\m_{. \al \n
\s}, \eqno{(49)}$$ $$ \h{i}^{\stackrel{\m}{-}} ||| {\n \s}  -
\h{i}^{\stackrel{\m}{-}} ||| {\s \n}   = \h{i}^\al {\*{L}}^\m_{.
\al \n \s},  \eqno{(50)}$$ $$ \h{i}^{\stackrel{\m}{0}} ||| {\n \s}
- \h{i}^{\stackrel{\m}{0}} ||| {\s \n}   = \h{i}^\al {\*{N}}^\m_{.
\al \n \s}, \eqno{(51)}$$

Note that every tensor with a hat is the parameterized replacement
of that without a hat. We can show that the tensors given by the
second method are more general than those obtained using the first
method. For example we can write, $$ W^\al_{. \m \n \s} = {\hat
{B}}^\al_{. \m \n \s} - b(b - 1) \g^\al_{. \m \be} \Lambda^\be_{.
\n \s}. \eqno{(52)}$$ We are going to discuss this result in the
next Section.\\

\section{Discussion and Conclusion}
Two main problems, concerning gravitational interactions, are now
well defined. The first is that quantization of gravity, using
conventional quantization schemes, is still a difficult task if
not impossible. The second, which may be a consequence of the
first, is that unification of gravity with other known
interactions is still beyond the reach of investigators. One way
to overcome these problems, my be in reexamining carefully the
geometrization scheme suggested and applied by Einstein in
constructing his theory of GR in free space. We are convinced that
this scheme is successful in constructing this theory, but
subsequent application of it, is the subject that needs a careful
examination.

It as well known that GR in free space is the most satisfactory
application of the geometrization scheme as shown in Section 2.
Recalling that the first criteria in this scheme is the choice of
an appropriate geometry for application; so in order to generalize
or modify this theory, one should first look for a geometric
structure more wider than the Riemannian one. For this reason
Einstein started two series of attempts, to construct unified
field theories, using two different geometries. We believe that
these attempts are incomplete rather than unsatisfactory. The
reasons are probably rooted in the geometric structures used. As
shown in Section 2, the geometric structures used are the
AP-geometry and the non-symmetric (NS) geometry. Table 1 shows how
many of the geometrization criteria, given in Section 2, are
satisfied by Einstein's two attempts, compared with GR in free
space as a standard theory of gravity. It is to be considered that
the version of GR, used for comparison in the present work, is
that for free space. It is the theory with less problems than the
version written for a material distribution. As stated before,
Einstein's two attempts are not successful.
 The reason may probably be the incompleteness of the geometries used.
We mean by a complete geometry, a geometry that satisfies the
following requirements:\\ i- All tensors, of different orders,
should be well defined in terms of the building blocks of the
geometry, especially those measuring curvature. Different
relations between these objects should be clarified.\\ ii- All
affine connections, admitted by the geometry, should be known and
well studied, with subsequent covariant derivatives defined using
these connections.\\ iii- Identities, especially those of the
differential type, are obtained for the geometry considered.\\ iv-
Different paths, admitted by the geometry, are obtained.\\ It is
obvious that Riemannian geometry satisfies all these requirements.
So, it could be classified as a complete geometry. But it is not
wide enough to modify or generalize GR. It is found to be just
sufficient to describe gravitational interactions, in free space,
with some limits concerning the distance from the source of the
field. The present work is a step towards making the AP-geometry
as complete as possible. This step is necessary before using this
geometry to overcome the problems mentioned above.

Table 2 displays geometrization attempts of some authors. The
AP-geometry is used, as a basic structure, in some of these
attempts [5], [8], [9], [10], [11], [12] and [13]. The NS-geometry
is used in another type of attempt [16]. In the rest of the
attempts, listed in the table, [14], [15] and [16] while a tetrad
defined in spaces with torsion is used, it is not clear whether
the AP- condition (2) is imposed on this tetrad. So, one cannot
decide whether these attempts were carried out using a version of
the AP-geometry or not.

In Section 3, ten reasons, for preferring AP-geometry for physical
applications, are given. However, the conventional version of this
geometry suffers from two main problems. The first is the
vanishing of its curvature tensor (28), which leads to the
conclusion that the geometry is flat. The second is that the path,
defined in this version, (26) does not represent physical
trajectories of any known particles. In the present work the two
problems are solved.\\ \underline{\it For The First Problem}: This
problem is solved on two levels. It is shown that the AP-space is
not flat. On the first level we have shown that the AP-space
admits non-vanishing curvature tensors (29)-(34), defined by the
second method given in Section 3. On the second level, by
parameterizing AP- connection (38), we were able to define a
non-vanishing curvature tensor (46) even by using the first method
given in Section 3. The curvature tensors, defined using the
second method, (49), (50) and (51) are also non-vanishing and more
general than those given in Section 3. Thus, the AP-geometry is no
longer flat. Even before defining the non-vanishing curvature
tensors admitted by the geometry, there were some evidences
indicating that this the AP-space is not flat. One of these
evidences is that GR can be written in this space with
satisfactory solutions (cf. [26]).\\ \underline{\it For The Second
Problem}: Also, this problem is solved on two levels. On the
first, we discovered that the AP-geometry admits other path
equations different from (26). These equations are given by the
set (27). This does not solve the problem completely, since
neither of these equations represents physical trajectories. But
this was a necessary step towards the solution, and to make the
geometry more complete. On the second level, we parameterized a
general path equation admitted by the geometry (45), which is used
now to represent physical trajectories in gravitational fields
[25], [27]. This step makes the AP- geometry more complete and
solves the second problem.

In modifying or generalizing GR in spaces with torsion (e.g.
[28]), the resulting theories are required to reduce to GR when
torsion vanishes. Such theories would encounter a problem in
conventional AP-geometry. The vanishing of the torsion in this
case would lead to the vanishing of all curvature tensors of the
structure [20]. So, the resulting GR would be a trivial one, and
gravity would no longer represented in such structures. These
theories are rendered unviable in the conventional version of the
AP-geometry. The situation now is different if such theories are
written in the parameterized AP-geometry. There is no need for the
torsion to vanish. It is sufficient, in order to reduce the theory
to GR, to switch the parameter off, which would correspond to some
physical condition. So, the problem of such theories is solved
using the parameterized version of the AP-geometry.

We can summarize the conclusion of the present work in the
following points.\\ 1- It is shown that the AP-geometry is not
flat since it admits a number of non-vanishing curvature
tensors.\\ 2- Paths admitted by the parameterized version of the
geometry can be used to study physical trajectories of test
particles.\\ 3- The parameterization carried out for the
AP-geometry make it more complete and more appropriate for
physical applications.\\ 4- The parameterized version of this
geometry is more wider than the Riemannian geometry and the
conventional AP-geometry. It can be reduced to the first upon
taking $b = 0$, $a = 1$ in (43), and to the second if we take $b =
1$, $a = 0$. \\ 5- The parameterized connection represents
simultaneously non-vanishing curvature and non-vanishing torsion.
This result is the contrary to what obtained by some authors
[6].\\ 6- The parameterized version of the AP-geometry is more
suitable for constructing field theories in spaces with torsion,
especially theories gauging gravity.\\

Finally, it is worth of mention that the geometrization criteria,
given in Section 2, are necessary but not sufficient to construct
geometric field theories.\\

A part of this work was carried out while the author was visiting
the H.E. Section of the Abdus Salam ICTP on January 1999. The
author would like to thank Professor Randjbar-Daemi, head of this
Section, for inviting him. Also, the author is grateful to
Professor Nutko,  and the organizing committee of the Regional
Conference on Mathematical Physics IX, for inviting him to give
this talk.
\newpage
\begin{center}
{\bf References}
\end{center}
{[1] P.D.B. Collins, A.D. Martin and E.J. Squires, Particle Physics and Cosmology, J.Wily and
    Sons (1989)}. \\
{[2] A. Einstein and L. Infeld, {\it Canadian J. Math.}, {\bf 1}
(1949) 209}. \\ {[3] A. Einstein, The Meaning of Relativity, 5th
ed., Princeton Univ. Press (1955)}. \\ {[4] A. Einstein, {\it
Sitz. der Preuss Akad der Wiss.}, {\bf 1} (1929) 1}. \\ {[5] K.
Hayashi and T. Shirafuji, {\it Phys. Rev.} {\bf D} {\bf 19} (1979)
12}. \\ {[6] V.C. de Andrade and J.G. Pereira, {\it Gen. Rel.
Grav.} {\bf 30} (1998) 263}. \\ {[7] V. Halavty, Geometry of
Einstein's Unified Field Theory, P. Noordhoff Ltd., N. Holland,
    (1957)}. \\
{[8] T. Levi-Civita, A Simplified Presentation of Einstein's Unified Field Equations,
     Blackie and Son Ltd. (1950)}.\\
{[9] F.I. Mikhail, Ph.D. Thesis, London University, London, UK,
1952}. \\ {[10] C. M\o ller, {\it Mat. Fys. Skr. Dan. Vid. Selsk.}
{\bf 39}, No. 1 (1961) 10}. \\ {[11] F.I. Mikhail, {\it IL Nuovo
Cimento}, {\bf X}, {\bf 32} (1964) 886}. \\ {[12] F.I. Mikhail,
M.I. Wanas, {\it Proc. Roy. Soc. Lond.} {\bf A} {\bf 356} (1977)
471}. \\ {[13] C. M\o ller, {\it Mat. Fys. Skr. Dan. Vid. Selsk.}
{\bf 39}, No. 13, (1978) 1}. \\ {[14] R. Hammond, {\it Gen. Rel.
Grav.}, {\bf 22} (1990) 451}. \\ {[15] R. Hammond, {\it Clas.
Quantum Grav.}, {\bf12} (1995) 279}. \\ {[16] J.W.   Moffat, {\it
J.Math. Phys.}, {\bf 36}(1995) 3722}. \\ {[17] R. Hammond, {\it
Gen. Rel. Grav.}, {\bf 30} (1998) 1801}. \\ {[18] H.P. Robertson,
{\it Ann. Math.}, {\bf 33} (1932) 496}.
\\ {[19] M.I. Wanas, M. Melek and M.E. Kahil, {\it Astrophys.
Space Sci.}, {\bf 228} (1995) 273}. \\ {[20] M.I. Wanas and M.
Melek, {\it Astrophys. Space Sci.}, {\bf 228} (1995) 277}. \\
{[21] R. Hammond, {\it Contemporary Phys.} {\bf 36} (1995) 103}.
\\ {[22] M.I. Wanas and M.E. Kahil, to appear in {\it Gen. Rel.
Grav.}, {\bf 31} (1999)}. \\ {[23] H.T. Nieh and M.Y.Yan, {\it
Ann. Phys.},{\bf 138} (1962) 237}. \\ {[24] F.I. Mikhail, {\it Ain
Shams Sci. Bull.}, {\bf 6} (1962) 87}. \\ {[25] M.I. Wanas,  {\it
Astrophys. Space Sci}, {\bf 258} (1998) 237}. \\ {[26] M.I. Wanas,
{\it Astron. Nachr.}, {\bf 311} (1990) 253}. \\ {[27] M.I. Wanas,
M. Melek and M.E. Kahil gr-qc/9812085 (1998)}. \\ {[28] F.W. Hehl
in "Spin, Torsion, Rotation and Supergravity" eds. P.G. Bergmann
and V. de Sabbata, Plenum Press (1980) 5}.\\
\end{document}